\documentclass[preprint,12pt,number]{elsarticle} %

\usepackage{graphicx} %
\usepackage{tcolorbox} %
\usepackage{tabularx} %
\usepackage{multirow}
\usepackage{booktabs}
\usepackage{pifont}%
\usepackage{tikz}
\usepackage{listings}
\usepackage{colortbl}
\usepackage{xcolor}
\usepackage[normalem]{ulem}
\usepackage[hyphens]{url}
\usepackage{acronym}
\usepackage{listings}
\usepackage{caption}
\usepackage{dirtree}
\usepackage{makecell}  %
\usepackage{hyperref}
\usepackage{amsmath}

\usepackage[scaled=.8]{beramono}
\lstdefinelanguage{diff}{
    morecomment=[f][\color{gray}]{@@},      %
    morecomment=[f][\color{red}]{-},        %
    morecomment=[f][\color{green!60!black}]{+}, %
    morecomment=[f][\color{blue}]{***},     %
    morekeywords={diff},
    sensitive=false,
}

\usepackage{xspace}
\usepackage[T1]{fontenc}  %
\usepackage{cleveref}  %
\usepackage{paralist}  %
\usepackage{array}  %
\usepackage{diagbox}
\usepackage{algorithm}
\usepackage{algpseudocode}

\algnewcommand\algorithmicinput{\textbf{Input:}}
\algnewcommand\Inputs[1]{\State\algorithmicinput~#1}
\algnewcommand\algorithmicoutput{\textbf{Output:}}
\algnewcommand\Outputs[1]{\State\algorithmicoutput~#1}

\def\BibTeX{{\rm B\kern-.05em{\sc i\kern-.025em b}\kern-.08em
    T\kern-.1667em\lower.7ex\hbox{E}\kern-.125emX}}

\newacro{sbom}[SBOM]{Software Bill of Materials}
\newacro{jvm}[JVM]{Java Virtual Machine}

\newcommand{\ssc}{Software Supply Chain\xspace}
\newcommand{\sbom}{\ac{sbom}\xspace}
\newcommand{\sboms}{\acp{sbom}\xspace}
\newcommand{\tool}{Classport\xspace}
\newcommand{\retriever}{Introspector\xspace}
\newcommand{\embedder}{Embedder\xspace}
\newcommand{\Retriever}{Introspector\xspace}
\newcommand{\Embedder}{Embedder\xspace}

\newcommand{\maven}{Maven\xspace}
\newcommand{\pom}{pom.xml\xspace}
\newcommand{\jvm}{\ac{jvm}\xspace}
\newcommand{\uberjar}{uber-JAR\xspace}

\newcommand{\uberjars}{uber-JARs\xspace}

\newcommand{\pdfbox}{PDFBox\xspace}
\newcommand{\ripper}{Certificate Ripper\xspace}
\newcommand{\mcs}{mcs\xspace}
\newcommand{\graphhopper}{GraphHopper\xspace}
\newcommand{\biojava}{BioJava\xspace}
\newcommand{\checkstyle}{Checkstyle\xspace}

\newcommand{\numberofprojects}{six\xspace}

\newcommand{\TODO}[1]{\textcolor{red}{#1}\GenericWarning{}{LaTeX Warning: TODO: #1}}\newcommand\todo\TODO

\newcommand{\added}[1]{\textcolor{black}{#1}}

\newcommand{\revision}[1]{\textcolor{black}{#1}}

\begin{document}

\begin{frontmatter}

\title{Classport: Designing Runtime Dependency Introspection for Java}

\author[imt]{Serena Cofano}
\ead{serena.cofano@imtlucca.it}

\author[kth]{Daniel Williams}
\ead{daniel\_williams@live.com}

\author[kth]{Aman Sharma}
\ead{amansha@kth.se}

\author[kth]{Martin Monperrus}
\ead{monperrus@kth.se}

\address[imt]{IMT School for Advanced Studies Lucca and University of Genoa, Genoa, Italy}
\address[kth]{KTH Royal Institute of Technology, Stockholm, Sweden}

\begin{abstract}
    Runtime introspection of dependencies, i.e., the ability to observe which dependencies are currently used during program execution, is fundamental for \ssc security.
    Yet, Java has no support for it. 
    We solve this problem with \tool, a blueprint and system that embeds dependency information into Java class files, enabling the retrieval of dependency information at runtime. We evaluate \tool on \numberofprojects real-world projects, demonstrating the feasibility in identifying dependencies at runtime. 
\end{abstract}

\begin{keyword}
    Software Supply Chain, Java, Maven
\end{keyword}

\end{frontmatter}

\pagestyle{plain}

\section{Introduction}\label{sec:intro}
Software systems keep increasing in size and complexity. 
To speed up implementation and satisfy the growing demand for features, developers often reuse existing  libraries~\cite{sonatypeReport}.
They leverage third-party open-source code, which is integrated into the main application as a dependency.
This collection of source code, dependencies, and build tools used to build a final software product is known as the \ssc~\cite{sonatypeReport}.

Using third-party dependencies is essential for productivity and reliability, but it also increases the attack surface~\cite{openssf2022}.
Security vulnerabilities that exist in a dependency are inherited by downstream software products, leading to potential compromises.
Notable \ssc incidents include the Log4Shell vulnerability~\cite{sonatype2021log4shell}, the Codecov compromise~\cite{sharma2021codecov}, and the xz catastrophe~\cite{przymus2025wolves}. 
Assessing and hardening the security of the \ssc requires increasing its so-called ``transparency'', i.e., the visibility on its components at all points in time of the software lifecycle~\cite{OkaforSok}.

One of the key tools to support transparency is the production of \sbom~\cite{cisaSBOM}.
An \sbom is an exhaustive record of the individual components included in a software system~\cite{ntiaSBOM}.
Maintainers can use this list as input for vulnerability assessment to detect malicious or vulnerable software packages. 
Most of the approaches to \sbom operate statically, by looking for dependencies in metadata files, such as manifests or lock files, only at build time~\cite{yu2024correctness,rabbi2024sbom, cofano2024sbom}.
This has two fundamental limitations.
First, it does not reflect the actual dependency usages at runtime, i.e., whether a dependency is actually loaded or used in production~\cite{BalliuChallenges}. 
Even if this does not present a direct risk for the application, it makes maintainers and developers waste a lot of time doing unnecessary work by assessing the security of unreachable dependencies, which slows down the response time to the actual security problems.
Second, static analysis of dependencies does not support the maintainers or developers to make runtime decisions based on the actually used dependencies.
Basically, the fundamental limitation of \sboms is that all dependency information is lost at runtime.

Runtime dependency introspection solves these limitations.
It is the ability of a piece of software to observe which components are executed during a program’s actual run, including component name and component version.
Runtime dependency introspection is the necessary building block for advanced security hardening.
First, it is crucial for identifying dependencies that are present in the application but never used at runtime. 
This simplifies vulnerability management by deprioritizing components that are never used and helps in reducing the attack surface through the removal of unused dependencies.
Second, runtime dependency introspection is required to be able to restrict execution privileges at runtime, by mapping the accessed resource to the corresponding dependency~\cite{amusuo2024ztdjavamitigatingsoftwaresupply}. 

The Java runtime is known for its good support of object-oriented introspection (aka class, field, and method reflection)~\cite{rigger2017introspection}. 
Yet, it is completely blind to dependencies: it has no native support for dependency introspection at runtime~\cite{xiao2025jbomaudit}. In Java, the information that uniquely identifies a dependency is present at build time, but it is not preserved at runtime.
Heuristics like guessing the origin of dependencies from package names have been implemented in previous works~\cite{amusuo2024ztdjavamitigatingsoftwaresupply}. However, they suffer from fundamental limitations, such as the requirement for dependencies to be compliant with fragile conventions to be correctly identified. Overall, the problem of runtime introspection of dependencies is an open one.
To the best of our knowledge, \textbf{there are currently no principled architecture for identifying which dependencies are actually executed at runtime in Java.}

In this paper, we propose \tool, a novel technique for introspecting dependencies at runtime in Java.
In essence, \tool brings dependency information to runtime: supplier, dependency name and version (group id, artifact id and version in Maven lingo).
\tool operates in two phases by:
\begin{inparaenum}
    \item embedding dependency information into Java binary artifacts through build-time instrumentation, and
    \item retrieving this information at runtime through dynamic instrumentation.
\end{inparaenum}

We evaluate \tool on \numberofprojects real-world open-source Java projects.
Our experimental protocol measures the ability to embed dependency information into the binary artifacts, and assesses the time and disk space overhead caused by the added information.
Then, we test \tool's ability to correctly identify the set of runtime dependencies using the test suite of the project.
We also measure the impact on the running application, incl. the runtime overhead.

Our results demonstrate the applicability of \tool to real-world applications.
\tool successfully embeds dependency information, accurately identifies runtime dependencies, preserves the functional behavior of the application, and introduces a low overhead. 

In summary, the main contributions of this paper are:
\begin{itemize}
    \item A novel approach to inspecting Java dependencies at runtime, based on Java annotations. To the best of our knowledge, it is the first ever solution for runtime dependency introspection in Java.
    \item \tool, a publicly available prototype implementing the blueprint: \tool embeds information into binary artifacts and provides a runtime agent to retrieve dependency information during execution~\cite{classport} (\href{https://github.com/chains-project/classport}{\revision{\path{https://github.com/chains-project/classport}}}).
    \item An evaluation of our novel technique on a set of \numberofprojects real-world applications, demonstrating the feasibility and applicability of our approach~\cite{classportExperiments} (\href{https://github.com/chains-project/classport-experiments}{\revision{\path{https://github.com/chains-project/classport-experiments}}}).
\end{itemize}

\section{Background}\label{sec:background}

This section provides background on how dependency management works in the Java ecosystem, with a focus on Maven and its limitations regarding runtime visibility of dependencies.

Dependency management by package managers and build systems varies across ecosystems, as they handle resolution, versioning, and transitive dependencies differently.
Java is one of the most popular programming languages~\cite{sosurvey2025}.
Its mature ecosystem of third-party libraries, distributed primarily through Maven, is critical in government services, financial services~\cite{SotoValeroMB22}, medical infrastructure, and enterprise software systems~\cite{MassacciP21}.
Maven Central hosts around 20 million artifacts depending on each other. This means that a single widely-used library can be transitively embedded in thousands of downstream projects, making Maven-based software supply chains both high-value and high-risk~\cite{MassacciP21,SotoValeroMB22}.
In the Java ecosystem, Maven is one of the available build systems~\cite{maven}.
Within Maven, developers declare dependencies in a file called \pom.
Each of these dependencies is downloaded as a binary artifact called a JAR~\cite{depMechanism}, and is uniquely identified by Group ID, Artifact ID, and Version~\cite{mavenPom}, this triple being called a GAV coordinate.
The Group ID represents the organization or project namespace, e.g., \textit{org.apache.commons}.
The Artifact ID specifies the module or library, e.g., \textit{commons-lang3}.
The Version distinguishes the same library across different releases, e.g., \textit{3.12.0}.

At build time, the dependencies of \pom are resolved, in order to list the GAV coordinates of both direct and transitive dependencies.
This allows Maven to download the exact version to use.
In cases of conflicts between multiple versions of the same transitive dependency, Maven's dependency resolution mechanism selects the version closest to the root of the dependency tree~\cite{depMechanism}.
The version visible in \pom is thus not necessarily the same one that will be used at runtime.
In Maven, dependency metadata such as the GAV of dependencies is only available at build time; this information is not preserved at runtime at all~\cite{xiao2025jbomaudit}.
The platform makes it impossible to identify the set of dependencies that are actually used at runtime, or to make any runtime decision depending on the dependencies.
\textbf{In other words, Java supports classes, methods, and fields introspection, but not dependency introspection}.

In this paper, we address this problem by proposing a blueprint architecture and a prototype for supporting dependency introspection in Java.

\section{Design of \tool}\label{tool}

\begin{figure}[h]
  \centering
  \includegraphics[width=0.9\textwidth]{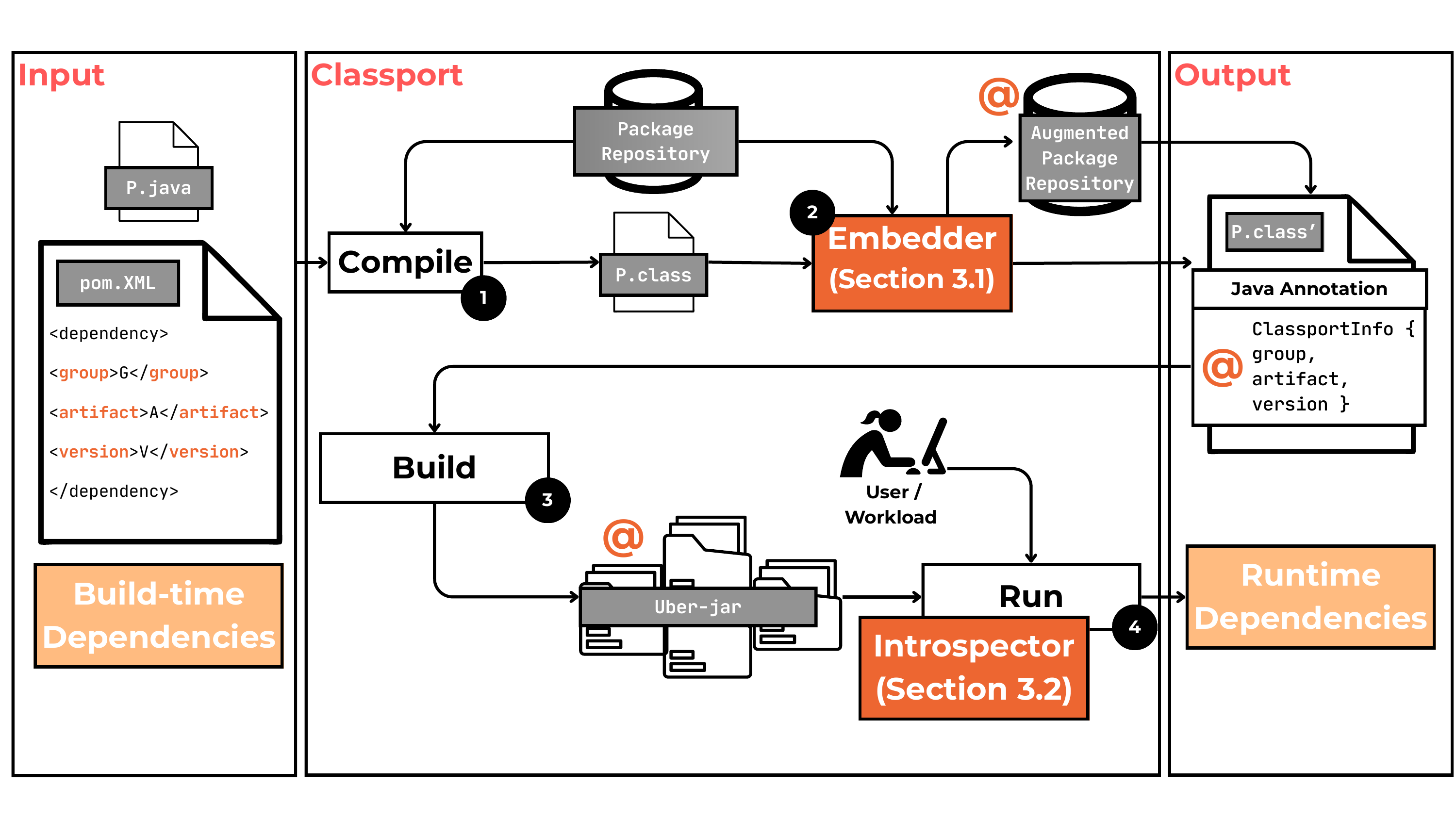}
  \caption{Overview of \tool, a novel system that enables runtime dependency introspection in Java.}
  \label{fig:overview}
\end{figure}

We propose \tool, a novel technique for runtime dependency introspection in Java.
Recall that in Java, dependency metadata is available at build time, but it is absent during execution.

We define \textbf{runtime dependencies} as dependencies that:
1) are fully specified within the binary to be executed, and
2) are available as variables from within the program.

We also provide an open-source implementation of \tool.
The main goal of \tool is to make \ssc information available at runtime, i.e. exposing the build-time Maven dependency metadata (Group ID, Artifact ID, and Version) that identifies each third-party component at runtime. \tool adds this information in the final executable artifact of a Java application, and client code can extract this information at runtime, during application execution.
\tool achieves this goal through its two key components: the \embedder and the \retriever.
The \embedder stores the information about dependencies in Java artifacts, while the \retriever extracts this embedded information with runtime introspection.

Figure \ref{fig:overview} presents an overview of \tool and its integration within the Maven lifecycle. 
\tool processes a target Java Maven project, which contains the source code and the information about build-time dependencies, which are listed in the \pom file.
The \embedder performs the \textit{embed} action (step 2), following the Maven compilation phase (step 1).
This produces the Java binary artifacts enriched with the dependency information, which are represented in the output box.
The $@$ symbol marks the artifacts that carry the embedded dependency information, such as the augmented package repository and the annotated class \textit{P.class'}.
This symbol follows Java syntax, in which every annotation starts with $@$, as in \texttt{@ClassportInfo}.
Next, the Maven build phase (step 3) packages these enriched artifacts into an \uberjar~\cite{mavenShade}, which contains the application together with all its resolved dependencies.
Finally, the \uberjar is run (step 4).
During this phase, the \retriever introspects the execution of the Java program and gives as output the runtime dependencies.
\tool is designed for applications packaged as \uberjars, which is the dominant deployment pattern for Maven applications (e.g., Spring Boot services and standalone command-line tools). An uber-JAR bundles all resolved dependencies at build time, so the build-time and runtime classpaths are identical by construction. 

The following two subsections detail the technical aspects of the two components of \tool.
Next, we present \tool with an end-to-end example in Section \ref{sec:end-to-end-example}.
Finally, we give details about the implementation of \tool, including tools and libraries used, in Section \ref{sec:implementation}.

\subsection{\Embedder}
The \textbf{\embedder} is the first key component of \tool, designed to bridge the gap between build-time and runtime visibility of dependency information. In Java Maven projects, dependency metadata is explicitly declared in the \pom and fully resolved during the build phase, but this information is not preserved in the final application artifacts. The \embedder addresses this limitation by capturing build-time dependency metadata and embedding it directly into the compiled artifacts, making it accessible at runtime.

The embedding process, illustrated in Figure~\ref{fig:overview}, proceeds in three stages.
First, the \embedder takes as input the compiled artifacts of a \maven project, i.e., the project’s class files and the JARs of the project’s dependencies.
The compilation phase, step 1 in Figure~\ref{fig:overview}, resolves the full transitive dependency graph of compile and runtime scopes  (test-scoped dependencies are excluded as they do not appear in the final \uberjar) and ensures that all dependency JARs are available locally.
Second, for each dependency JAR, the \embedder locates the local version that was downloaded and processes each file within the JAR.
Depending on the resource it encounters, this processing involves different actions.
For class files within a dependency JAR, i.e., compiled Java source files with \texttt{.class} extension, the \embedder embeds dependency metadata, i.e., its group, artifact, and version, using a bytecode transformation.
If there are duplicate classes in the dependency JAR, the \embedder will embed the metadata for both classes.
However, the JVM will only load one of them, so only one of the metadata will be visible at runtime.
The same process is applied to the application classes, such as \textit{P.class} in  Figure~\ref{fig:overview}, annotated with the project module information.

When the \embedder encounters manifest files with a dependency JAR, such as \textit{MANIFEST.MF} files, it removes signature-related entries, e.g., digest attributes, and copies the remaining content. 
This is necessary because modifying the contents of a signed JAR, such as adding metadata to class files, invalidates its signature.
However, \tool is designed to be used by application developers and they can resign the JAR with their keys after the embedding process.
\revision{This workflow is well-established and is also required by other widely-used bytecode-transformation tools such as JaCoCo~\cite{jacoco}.}
For instance, consider a project $P$ authored by an application developer $A_1$ that depends on a JAR $D$ signed by a third-party library developer $A_2$.
During embedding, we strip $A_2$'s signature from $D$ so that we can modify the classes, and the embedded dependency is then  re-signed by $A_1$ once the process completes, as in any vendoring process.
The resulting \uberjar is distributed by $A_1$, who is responsible for the packaged software.
The \embedder ignores signature files and copies all other files without modification. We refrain from altering non-class files.
If these other files are used, they must be referenced by class files that are themselves embedded.
If they are not used, the Java Virtual Machine will not load them, and hence we don't need to get provenance later.
After processing each dependency JAR, it repackages it into a new JAR. It saves it in a directory, which we refer to as the augmented package repository, as shown in ~\autoref{fig:overview}.
It is noteworthy that the embedding process is not affected by shading (i.e., package renaming performed by the Maven Shade Plugin), because the \embedder injects the correct GAV metadata before any renaming takes place.

Third, as shown in step 3 of ~\autoref{fig:overview}, the augmented \uberjar with embedded software supply chain information is pushed where appropriate.

\begin{algorithm}[!tb]
\caption{Algorithm for the \embedder process.}
\label{alg:embedder}
\small
\begin{algorithmic}[1]
\Inputs{\textit{pom:} The Maven project along with its dependencies}
\Outputs{\textit{uberjar:} Augmented \uberjar with embedded dependency metadata}

\State $deps \gets \textproc{resolve\_dependencies}(pom)$

\For{each dependency $d$ in $deps$}
  \State $jar \gets \textproc{fetch\_jar}(d)$
  \State $entries \gets \textproc{open\_jar}(jar)$
  \State $out \gets []$
  \For{each entry $e$ in $entries$}
    \If{$\textproc{is\_class\_file}(e)$}
      \Comment{magic bytes \texttt{0xCAFEBABE}}
      \State $e \gets \textproc{inject\_annotation}(e, d.\mathit{GAV})$
      \Comment{Bytecode transformation}
      \State $out.\mathit{add}(e)$
    \ElsIf{$\textproc{is\_manifest}(e)$}
      \Comment{\texttt{MANIFEST.MF}}
      \State $e \gets \textproc{strip\_signature\_entries}(e)$
      \State $out.\mathit{add}(e)$
    \ElsIf{$\textproc{is\_signature\_file}(e)$}
      \Comment{\texttt{.SF}, \texttt{.RSA}, \texttt{.DSA}}
      \State \textbf{skip}
    \Else
      \State $out.\mathit{add}(e)$
      \Comment{copy unchanged}
    \EndIf
  \EndFor
  \State $repo.\mathit{add}(\textproc{repackage\_jar}(out, d))$
\EndFor

\State \textbf{invoke} Maven \texttt{package} phase to assemble final \uberjar
\State \Return $uberjar$

\end{algorithmic}
\end{algorithm}

The summary of these steps is presented in ~\autoref{alg:embedder}.

\subsubsection{\Embedder as a Maven plugin} 
\revision{We use a Maven plugin to implement the \embedder~\cite{mavenPlugins}.}
A Maven plugin is a reusable and highly configurable component that adds specific tasks or goals to Maven's build process.
In particular, it is possible to decide in which build phase to execute it by configuring a Maven plugin goal.

Within \tool, we create the custom \textit{embed} goal. 
We configure it to fulfill the following needs of the plugin: 
\begin{inparaenum}
    \item to have project's class files available locally. This is fundamental because they have to be embedded.
    \item To have dependencies' class files. Dependencies must be resolved to also have both direct and transitive ones.
\end{inparaenum}

The first point is achieved by compiling the target project before executing the plugin.
It guarantees that the class files for the project are available for processing.
\revision{The second point is fulfilled by ensuring that the resolution of dependencies happens before the embedding process~\cite{mavenResolutionScope}.}
\added{The embedding and then packaging process happens during a single Maven \texttt{package} phase.}

\subsubsection{Information embedded as Java Annotations}
The primary objective of the tool is to introspect dependency information during execution, so it is crucial to embed data in a manner that allows for retrieval at runtime. 
We use Java Annotations to add such information into the binary class files. We use bytecode transformation 
for that task. %
In each class, we create a custom annotation \textit{ClassportInfo}, from the annotation type shown in Listing~\ref{lst:annotation}.

\begin{lstlisting}[language=Java, caption={The Classport Java annotation for storing dependency information within the code. The annotation is available at runtime through introspection.}, label={lst:annotation},float]
@Retention(RetentionPolicy.RUNTIME) 
public @interface ClassportInfo { 
    String group(); // G
    String artefact(); // A
    String version(); // V    
}
\end{lstlisting}

The annotation contains the GAV coordinate, i.e., group, artifact, and version of the dependency to which the class belongs.
The Retention annotation~\cite{retention}, \texttt{@Retention(RetentionPolicy.RUNTIME)}, is used to make the \texttt{ClassportInfo} annotation visible at runtime.
After the injection of \ssc information by the \embedder, the target application is packaged as a \uberjar.

\subsection{\Retriever}
As illustrated in phase 4 of Figure~\ref{fig:overview}, the \retriever instruments the running application, reading embedded annotations from classes at load time and intercepting method invocations to record which dependency each invoked method belongs to.
This process allows \tool to determine which dependencies are actually used during execution, providing a precise view of the runtime dependency set.

\subsubsection{\Retriever as Java agent using the Instrumentation API}
The \retriever is a Java agent attached to the JVM.
The goal of the \retriever is to add code that extracts the dependency annotations.
To achieve this, we instrument the methods of the application under study using code transformation.
Each time an instrumented method is executed, the extraction logic is triggered, storing the current dependency GAV in a set.
To maintain a low overhead, not every method is instrumented.
The \retriever intercepts each class at load time, before it is first used. It reads the \texttt{@ClassportInfo} annotation and checks whether the corresponding GAV is already present in an in-memory set. If not, it rewrites the class bytecode to insert a small recording snippet at each method entry. If the GAV is already recorded, the class is left uninstrumented. Each class is therefore loaded exactly once, and no previously instrumented bytecode is ever modified again.
This allows us 1) to drastically reduce the overhead and 2) returns the exact set of dependencies used during execution.
Finally, after execution of the workload, \Retriever writes the resulting set of dependency GAVs into a CSV file.

While the embedded \texttt{@ClassportInfo} annotations can be read via plain reflection, a Java-agent approach is preferable for two reasons.
First, no application modification is required.
Reflection can read annotations at runtime but would require adding explicit calls such as \texttt{\path{clazz.getAnnotation(}} \texttt{\path{ClassportInfo.class)}} to the application source, with the developer knowing in advance which classes to inspect.
The \retriever attaches to the JVM at startup and transparently intercepts any application without source-level changes.
Second, the \retriever performs execution tracking rather than static enumeration.
A dependency is recorded only when one of its methods is actually executed, identifying the subset of dependencies that are genuinely reachable during a given workload.
Plain reflection cannot construct this set because reflection APIs hard code the set of classes to inspect.

\subsection{End-to-end Example}\label{sec:end-to-end-example}

\revision{Listing~\ref{lst:diff} shows the changes \tool makes on the original bytecode of the class \texttt{ExtractText} in the \texttt{PDFBox} project~\cite{pdfbox}.}
In particular, it shows the resulting bytecode modifications after running the \embedder. 
\tool adds an attribute in the attribute section (lines 12 to 17), which is the representation of the annotation.
The keyword \texttt{RuntimeVisibleAnnotations} (line 12) indicates the retention policy that has been set, and that the annotation is visible at runtime.
The remaining part of the section represents the key-value pairs of the annotation, i.e., the Group, Version, and Artifact.
Finally, the constant pool section (lines 4-9) is consequently updated with the added values.

Listing~\ref{lst:pdfbox-deps} shows the CSV output content of the \retriever. 
It reports the list of dependencies used during the execution of the application.
Specifically, the Group, Artifact, and Version are reported for each dependency.

\begin{lstlisting}[language=diff, caption={Example of resulting Java bytecode after embedding process.}, label={lst:diff},float]
Classfile /.../ExtractText.class
     ...
   Constant pool:
     ...
    #89 = Utf8               SourceFile
    #90 = Utf8               ExtractText.java
+   #91 = Utf8               Lio/github/.../ClassportInfo;
+   #92 = Utf8               group
+   #93 = Utf8               org.apache.pdfbox
     ...
   Attributes:
+    RuntimeVisibleAnnotations:
+      @ClassportInfo(
+         group = "org.apache.pdfbox",
+         version = "3.0.4",
+         artefact = "pdfbox-tools",
+      )
\end{lstlisting}

\begin{lstlisting}[language={},caption={\added{Result of runtime dependency introspection for PDFBox workload execution. Each row is a triple of Group, Artifact, and Version.}},label={lst:pdfbox-deps},float]
commons-io,commons-io,2.18.0
commons-logging,commons-logging,1.3.4
info.picocli,picocli,4.7.6
org.apache.pdfbox,fontbox,3.0.4
org.apache.pdfbox,pdfbox,3.0.4
org.apache.pdfbox,pdfbox-examples,3.0.4
org.apache.pdfbox,pdfbox-io,3.0.4
org.apache.pdfbox,pdfbox-tools,3.0.4
org.apache.pdfbox,preflight,3.0.4
org.apache.pdfbox,xmpbox,3.0.4
org.bouncycastle,bcpkix-jdk18on,1.80
org.bouncycastle,bcprov-jdk18on,1.80
org.bouncycastle,bcutil-jdk18on,1.80
\end{lstlisting}

\subsection{Usage}
\label{sec:usage}
The application developer is expected to run the packaging phase in Maven with our plugin configured.
Our Maven plugin automatically embeds the dependency metadata into the compiled classes, producing an enriched JAR that can be published and distributed to package registries.
This enriched JAR enables two modes of dependency introspection.
Any tool can statically scan its \texttt{@ClassportInfo} annotations to enumerate all bundled dependencies without executing the application.
Users of such JARs also get the novel ability to introspect the embedded dependency tree at runtime using the \retriever component, which identifies the subset of dependencies that are genuinely loaded and executed during a given workload.
Since \tool is automatically integrated within the packaging phase, keeping the embedded dependency information up-to-date is fully automated in the release process.

\subsection{Implementation}\label{sec:implementation}
\tool is implemented in Java.
\revision{It is made of two modules: the \embedder uses the ASM~\cite{asm} library to manipulate the bytecode in the class files, while the \retriever is a Java agent that uses the Instrumentation API~\cite{instrAPI}.}
The implementation is open-source and available on GitHub~\cite{classport}.

To use Classport, application developers declare the \embedder as a Maven plugin in the \texttt{<build><plugins>} section of their \pom, so it does not appear on the runtime classpath of the downstream user.
The \retriever, however, must be declared as a runtime dependency by those who wish to introspect dependency information at runtime, as it needs to read the \texttt{@ClassportInfo} annotations from loaded classes.

\section{Experimental Methodology}\label{sec:methodology}
This section outlines the methodology for our study evaluating the implementation of \tool. 
First, it presents the research questions that guide our investigation and information about the dataset.
Then, it provides a detailed explanation of how we address these research questions.
Lastly, it shows the overall experimental setup. 
\subsection{Research Questions}
We evaluate \tool to demonstrate its effectiveness in introspecting runtime dependencies in real-world applications.
First, we assess its capability to embed dependency information into Java binary artifacts. 
Then, we test its behaviour during runtime execution and its ability to retrieve the actual set of running dependencies.
Specifically, we investigate the following research questions:

\textbf{RQ1}: To what extent can \tool effectively embed dependencies into Java binary artifacts?

\textbf{RQ2}: To what extent does \tool support runtime inspection of dependencies?

\begin{table}[]
  \centering
  \caption{Study subjects considered in our experiments.}
  \label{tab:projects}
  \resizebox{1\columnwidth}{!}{
  \begin{tabular}{|l|c|c|c|}
    \hline
    \textbf{Project} & \textbf{\added{\#Module}} & \textbf{Version} & \textbf{Transitive Dependencies} \\ \hline
    \href{https://github.com/apache/pdfbox}{\pdfbox} 
    & \added{13}
    & \href{https://github.com/apache/pdfbox/tree/3.0.4}{3.0.4} 
    & \added{19} \\ \hline

    \href{https://github.com/Hakky54/certificate-ripper}{\ripper} 
    & \added{1}
    & \href{https://github.com/Hakky54/certificate-ripper/tree/2.4.1}{2.4.1} 
    & 5 \\ \hline %

    \href{https://github.com/mthmulders/mcs}{\mcs} 
    & \added{1}
    & \href{https://github.com/mthmulders/mcs/tree/v0.7.3}{0.7.3} 
    & \added{5} \\ \hline

    \added{\href{https://github.com/biojava/biojava}{BioJava}} 
    & \added{14}
    & \added{\href{https://github.com/biojava/biojava/releases/tag/biojava-7.2.4}{7.2.4}} 
    & \added{50} \\ \hline

    \href{https://github.com/checkstyle/checkstyle}{checkstyle} 
    & \added{1}
    & \href{https://github.com/checkstyle/checkstyle/tree/checkstyle-10.23.0}{10.23.0} 
    & \added{37} \\ \hline

    \added{\href{https://github.com/graphhopper/graphhopper}{GraphHopper}} 
    & \added{11}
    & \added{\href{https://github.com/graphhopper/graphhopper/releases/tag/11.0}{11.0}} 
    & \added{172} \\ \hline
  \end{tabular}
  }
\end{table}

\subsection{Dataset}
\label{sec:dataset}
We aim at evaluating \tool with a dataset of real-world, maintained, open-source Java projects, which use Maven and span a range of application domains, dependency sizes, and usage scenarios.
Projects are manually selected according to the following criteria: (1)~the project is open-source and actively maintained; (2)~it uses Maven as its build system; (3)~the tagged release compiles and its test suite passes; and (4)~projects were chosen to maximize diversity in dependency scale and module count, following the selection methodology of prior runtime-analysis studies~\cite{sharma2024sbomexecounteringdynamiccode,10795273,10677447}.
\added{Our dataset includes \numberofprojects projects, with projects varying in complexity, and includes both simple tools such as \mcs and large frameworks such as \checkstyle. The most popular repository is \checkstyle, which has around 9K stars on GitHub.}
\added{Table~\ref{tab:projects} lists each considered Maven project along with its number of modules, version, and number of dependencies.
The modules refer to the Maven modules (also called subprojects) of the project.
For example, \pdfbox has 13 modules and \texttt{app} and \texttt{core} are two of them.}
\added{The considerred version is the tagged release of the project.}
The number of transitive dependencies is calculated directly by Maven.
\added{We compile and embed all projects for RQ1 and analyze the resulting artifact. For RQ2, we run the \retriever while executing the tests of the projects in the dataset.}

\subsection{Methodology of RQ1}

The first research question aims to evaluate the \embedder.
We assess the dependency completeness, class completeness, and the performance overhead of embedding dependency metadata into Java binary artifacts using \tool.
This is done by statically analyzing the JAR file produced by \tool.

\textbf{Dependency Completeness}
This property ensures that \tool embeds the \uberjar with dependency metadata of all dependencies.
To assess this, we first run the \embedder on the dataset, generating the annotated \uberjars.
We only consider dependencies that are needed for the application's execution, excluding test-only dependencies, as they do not end up in the \uberjar.
Then, we analyze the \uberjars to read the annotations inside the binary artifacts.
\added{We need to only consider the \uberjar because it contains all the dependencies (as opposed to 
a standard JAR which only contains the classes of the project and not the dependencies).}
We consider the embedding process complete in terms of dependency metadata if the dependency metadata, i.e., each Group ID, Artifact ID, and Version, matches the ground truth retrieved by running the Maven command \texttt{mvn dependency:list} on the project under test. 

\textbf{Class Completeness}
We assess completeness by verifying that the annotation is embedded in each class file of every dependency in the project.
Although the dependency can have files other than class files, such as build manifests and various configuration files~\cite{sharma_causes_2025}, we only consider class files for annotation as they are the only ones that can be executed.
To verify completeness, we first run the \embedder on the study subjects.
Then, we analyse the \uberjar, looking inside each class file.
If all Java class files are annotated with the \tool annotation, the embedding process is complete in terms of class metadata.

\textbf{Build Time Overhead}
Build time overhead is measured by comparing the time taken to build the \uberjar with and without the \embedder.
In detail, we run package phase of the Maven build process with and without the \embedder and measure the time taken for each run using the Unix \texttt{time} command.
\added{We do the runs 10 times to accommodate for variance and we report the median of the overhead incurred.}

\textbf{Disk Overhead}
Disk overhead is the additional space required to store the embedded dependency metadata.
To measure the disk overhead, we compare the sizes of the \uberjar before and after embedding. %
Specifically, we run the same Maven command as in `Build Time Overhead' to build the \uberjar with and without the \embedder and measure the disk usage for each run using the Unix \texttt{stat} command. We compute the difference in size of the JAR between the two runs and report the percentage of overhead.

\subsection{Methodology of RQ2}
The second research question evaluates the behavior of \tool at runtime, in particular, its capability to introspect runtime dependencies. 
\added{The objective is to show that 1) \tool can identify which dependencies are currently running and 2) the target application's runtime behavior is not affected by \tool.}  

To evaluate the \retriever, it is necessary to first run the \embedder, to ensure that the embedding process is completed and also that all the dependency metadata is embedded into the \uberjar. 
This means that we assume a correct \embedder, as validated by RQ1, and we now assess the \retriever, which is the component dedicated to the runtime introspection.

\textbf{Runtime Dependency}\label{sub:rq2-runtime-introspection}
\added{We report the dependencies detected during workload execution.}
\added{As workload, we run the test suite of each study subject in \autoref{tab:projects} on the embedded project and record the runtime dependencies it triggers.}
This produces a list of runtime dependencies observed by \tool.
\added{The observed list should contains all embedded dependencies which are exercised at runtime.}

\textbf{Overhead}
\added{To measure the runtime impact of \tool, we measure the time taken to execute the test suite with and without the \tool agent.
The overhead percentage is computed as $\text{overhead}~(\%) = (T_{\text{with agent}} - T_{\text{without agent}}) / T_{\text{without agent}} \times 100$.
The absolute time difference (in seconds) is reported in parentheses next to the percentage in Table~\ref{tab:results-rq2}.
We also do this measurement 10 times and report the median of the overhead incurred.}

\textbf{Functional correctness} For validating that \tool does not break correctness, we run the test suite of the target application and monitor for runtime errors or test failures caused by the agent.

\subsection{General Setup}
We run the experiments on a server with an Intel i9-10980XE CPU (36 cores), 125 GB RAM, using Java 17.0.15 and Maven 3.8.7.
The dataset and experiment scripts are publicly available on GitHub~\cite{classportExperiments}.

\section{Experimental Results}\label{results}
This section presents the results of the experiments as detailed in \autoref{sec:methodology}.

\subsection{Results for RQ1}

\begin{table}[]
  \centering
  \caption{RQ1 – Static Completeness and Overhead of Embedding Process}
  \label{tab:results-rq1}
  \resizebox{1\columnwidth}{!}{
  \begin{tabular}{|l|ll|ll|}
  \hline
  \multirow{2}{*}{\textbf{Project}} & \multicolumn{2}{c|}{\textbf{Completeness}} & \multicolumn{2}{c|}{\textbf{Overhead}} \\ \cline{2-5} 
                                    & \multicolumn{1}{c|}{\textbf{Dependencies}} & \multicolumn{1}{c|}{\textbf{Classes}} & \multicolumn{1}{c|}{\textbf{Time (s)}} & \multicolumn{1}{c|}{\textbf{Space (MB)}} \\ \hline
  \pdfbox-app (app)                        & \multicolumn{1}{l|}{12/12}        & 7,914/7,914  & \multicolumn{1}{l|}{\added{11.52\%(2.4)}}  &  \added{13.82\%(1.9)} \\ \hline
  Certificate-ripper (itself)                & \multicolumn{1}{l|}{4/5}          & 426/426      & \multicolumn{1}{l|}{\added{6.80\%(0.2)}}   & \added{13.17\%(0.1)} \\ \hline
  \mcs (itself)                               & \multicolumn{1}{l|}{4/4}          & 557/557      & \multicolumn{1}{l|}{\added{7.62\%(0.3)}}   & \added{9.88\%(0.1)} \\ \hline
  \biojava (biojava-aa-prop)                             & \multicolumn{1}{l|}{\added{31/32}}          & \added{7,629/7,629}  & \multicolumn{1}{l|}{\added{7.17\%(1.2)}}   & \added{17.19\%(2.4)} \\ \hline
  \checkstyle (itself)                        & \multicolumn{1}{l|}{\added{34/37}}        & 10,691/10,691 & \multicolumn{1}{l|}{\added{11.38\%(2.7)}}  & \added{17.20\%(3.4)} \\ \hline
  \graphhopper (graphhoper-web)                             & \multicolumn{1}{l|}{\added{122/166}}          & \added{20,965/20,965}     & \multicolumn{1}{l|}{\added{16.80\%(6.5)}}   & \added{12.96\%(6.1)} \\ \hline
  \end{tabular}
  }
\end{table}

The results of RQ1 are summarized in Table~\ref{tab:results-rq1}.
It presents the effectiveness of \tool in embedding dependency metadata into Java class files across the open-source projects tested.
Each row of the table represents the results of a project on our dataset.
\added{The first column lists the projects and their executable \uberjar  from our dataset described in \autoref{sec:dataset}.}
\added{The second column reports the completeness of the embedding process considering two aspects.}
\added{The first aspect is the number of dependencies embedded compared to the total number of dependencies in the project, and the second one is the number of class files containing the \tool annotation, compared to the total number of class files in the original \uberjar}.
Then, the table reports the overhead introduced during the build phase by the \embedder.
\added{We report the embedding time overhead, i.e., the time added by the embedding process to the build, and the space overhead, i.e., the size of the \uberjar after embedding.
We also show the absolute increase in paranthesis.}

Consider the first row of the table, which corresponds to the \pdfbox project.
\tool successfully embeds all 12 dependencies of \pdfbox-app, which is the total number of dependencies required by the application.
The output \uberjar contains 7,914 class files, and all of them are annotated with the dependency metadata.
\added{For this project, the embedding process introduces a time overhead of 11.52\% and a space overhead of 13.82\% in the final \uberjar.}

\added{There are two projects, \pdfbox and \mcs, which are completely embedded, while the other projects are partially embedded.}
\added{There are three reasons why an \uberjar does not have all the dependencies as declared in the \pom file.}
\added{The first one is the case where a dependency JAR does not contain any class files.
It may be empty, which happens for dependency \texttt{com.google.guava:listenablefuture:9999.0-empty-to-avoid-conflict\\-with-guava}, or it may contain files such as JavaScript, Kotlin metadata, or other non-class files.
\checkstyle and \graphhopper are examples of projects that have dependencies that do not contain any class files.}
\added{The second reason is where a dependency is only necessary for compilation and not for runtime, which is called a \textit{provided} dependency in Maven.
This dependency is not included in the \uberjar by default.
\ripper and \checkstyle are examples of projects that have dependencies that are only necessary for compilation and not for runtime.}
\added{Finally, the third reason is where classes of a dependency are shadowed by the other.
In this case, \biojava has the dependency \texttt{javax.xml.bind:jaxb-api} which shadows \path{jakarta.xml.bind:jakarta.xml.bind-api}.}
\added{Note that shadowing is not a consequence of the \embedder, but a consequence of how Maven packages the \uberjar~\cite{reyes_mavenhijack_2025}.}
\added{In all three cases, the classes from those dependencies are not executed.
Recall that \tool is designed to detect dependencies that are executed during runtime.
Therefore, these dependencies are rightfully not detected by \tool.}

The embedding process introduces a moderate performance cost during the build phase.
\added{On an absolute scale, the embedding time overhead ranges from 0.2s to 6.5s which we claim is acceptable in a build.}
\added{The space overhead varies from 0.1MB to 6.1MB which is considered acceptable given the introduction of a novel feature: visibility into runtime dependencies.}
The space overhead depends upon the number of dependencies and identifier names in the core binary content.
\ripper and \mcs have about the same number of classes and the exact same number of dependencies embedded, yet the overhead is larger for \ripper due to longer identifier names.

\begin{tcolorbox}[colback=gray!10, colframe=black!50]
Answer to RQ1: Our experiments show that \tool completely embeds dependency metadata into Java class files. Our benchmark of six diverse Maven projects demonstrates that \tool works with  real-world and complex software.
Build time and space overhead are an acceptable trade-off for getting a unique feature absent from the Java stack: runtime visibility into  dependencies.
\end{tcolorbox}

\subsection{Results for RQ2}
\begin{table}[]
  \centering
  \caption{RQ2 – List of Runtime Dependencies and Overhead of Introspector.}
  \label{tab:results-rq2}
  \resizebox{1\columnwidth}{!}{
  \begin{tabular}{|l|l|l|l|}
  \hline
  \textbf{Project} & \makecell{\textbf{Runtime} \\ \textbf{Dependencies}} & \textbf{Overhead (s)} & \makecell{\textbf{Functional} \\ \textbf{Correctness}} \\ \hline
  \pdfbox                       & \added{13}           & \added{4.11\%(1.9)} & \ding{51} \\ \hline
  Certificate-ripper               & 4           & \added{2.95\%(0.1)}  & \ding{51} \\ \hline
  \mcs                              & \added{4}           & \added{5.96\%(0.3)}   & \ding{51} \\ \hline
  \biojava                            & \added{23}           & \added{8.54\%(56.8)}   & \ding{51} \\ \hline
  \checkstyle                       & 22           & \added{5.28\%(8.8)} & \ding{51} \\ \hline
  \graphhopper                          & \added{104}           & \added{5.22\%(11.0)}   & \ding{51} \\ \hline
  \end{tabular}
  }
\end{table}

We now assess to what extent we are able to obtain the dependency information during execution. 
Table~\ref{tab:results-rq2} summarizes the effectiveness of \tool in detecting dependencies at runtime across the evaluated applications.
\added{The table is divided into three parts: the first part reports the results regarding the detected runtime dependencies, while the second shows the runtime overhead (as opposed to the build overhead in RQ1), and the third one reports the functional correctness.}

\added{\emph{Detected Runtime Dependencies} This is the number of dependencies detected by \tool during the execution of the workload.
For \ripper and \mcs, the workload covers the entire set of embedded dependencies. 
For the other applications, not all embedded dependencies are detected at runtime.
On manual inspection, we found that this is due to two reasons.
First, some dependencies are not exercised by the workload or are bloated.
Such as the dependency \path{io.dropwizard.logback:logback-throttling-appender:1.5.1} of \graphhopper is not used by the workload.
Second, some dependencies have classes that are loaded by the \jvm but not executed.
This happens because \jvm loads classes to check for type safety, declared exceptions, and extracting metadata from annotations.
For example, the dependency \path{com.google.errorprone:error_prone_annotations:2.36.0} only has annotations.
Annotations in Java are not executed as they don't have any methods.
For these two reasons, \tool does not detect these dependencies because it only considers dependencies that are executed during runtime.}

\emph{Overhead.} The overhead during execution of is low for all the study subjects.
It varies from 2.95\% to 8.54\%.
The overhead is due to the fact that \tool needs to instrument methods invocation with an agent in order to extract the dependency information.
Note that this runtime overhead is entirely incurred by the \retriever, and not by the \embedder and thus, it is different from the build overhead reported in RQ1.

\added{\emph{Functional correctness} \tool does not break any functionality as the entire test suite runs with the same result with or without introspection.
This means that the bytecode instrumentation done by \tool does not introduce bugs on the target application.}

\begin{tcolorbox}[colback=gray!10, colframe=black!50]
\added{Answer to RQ2: \tool can successfully introspect dependencies at runtime, and allows applications to identify which dependencies are being currently executed. \tool does not affect application behavior. The execution overhead is low for all study subjects. To our knowledge, \tool is the first system in the Java ecosystem to support dependency introspection at runtime.} 
\end{tcolorbox}

\section{Use Cases of \tool}\label{sec:usecase}
This section describes the possible benefit of having dependency information available at runtime in Java.
\tool provides the foundational capability of knowing which dependencies are being executed, enabling the security use cases described in this section.

\subsection{Runtime Permissions per Dependency}
Limiting what resources (e.g., network, filesystem) software components can access at runtime is relevant for security.
In the context of \ssc security, a promising strategy is to assign permissions per dependency, ensuring that each third-party library is granted only the privileges it strictly requires.
This permission enforcement per dependency requires precise identification of dependencies at runtime~\cite{amusuo2024ztdjavamitigatingsoftwaresupply, cesarano2025goleashmitigatinggolangsoftware}.

\tool enables this by exposing stable GAV coordinates for each dependency executing at runtime.
For example, a policy file can grant network-access permission only to \texttt{\path{org.apache.httpcomponents.client5:httpclient5:5.6}}.
When the \retriever detects at runtime that a method from this dependency is actually executed, a permission manager enforces the declared policy based on the dependency's GAV identity.
This is consistent with advisory databases such as NVD and OSV, which maintain vulnerability records keyed on GAV coordinates, making GAV the natural anchor for security policy enforcement.
Approaches that instead infer identity from fully qualified package names, such as that of Amusuo et al.~\cite{amusuo2024ztdjavamitigatingsoftwaresupply}, are unreliable in practice.
The classes of \texttt{httpclient5} reside under \texttt{org.apache.hc.*}, but after shading they may be relocated to \texttt{com.example.shaded.hc.*}, so a permission mapped to the original package name would no longer match the executing classes, leaving the policy silently unenforced.
GAV coordinates are injected by the \embedder before any shading takes place, so they remain correct regardless of how packages are subsequently renamed.

\subsection{Vulnerability Detection at Runtime}
Vulnerability detection tools aim to identify known vulnerabilities in third-party libraries.
Identifying dependencies that are actually used at runtime is fundamental to avoid false positives or false negatives in vulnerability assessment~\cite{williams2025directions}.

\tool addresses this gap by making it possible to observe exactly which dependencies are executed during a program’s run. By embedding GAV metadata into class files and retrieving it dynamically, \tool enables runtime-aware vulnerability detection tools that focus only on actually used components.
Consider CVE-2022-23302~\cite{ghsa2022log4j1jmssink}, which affects \texttt{log4j:log4j} up to and including version 1.2.17.
The vulnerability is a deserialization flaw in the JMSSink component, allowing remote code execution.
Log4j 1.x has reached end-of-life, and no patched release exists within the 1.x line.
Remediation requires migrating to Log4j 2.x, a complete library replacement.
The Maven artifact identifier, package names, and configuration format all change between the two versions, so every logging call and configuration file in the application must be updated.
For a large codebase, this migration is a significant effort.
A static SCA scan will flag every service whose transitive dependency graph includes \texttt{log4j:log4j:1.2.17}, generating a uniform critical alert across all services.
With \tool, a security team can integrate with the \retriever to query each deployed service while it runs in production.
\revision{Services where \texttt{log4j:log4j:1.2.17} appears in the \retriever’s output, a sample of which is shown in Listing~\ref{lst:pdfbox-deps}, are confirmed as actively executing the vulnerable library and become the first migration targets.}
Services where the dependency does not appear can be assigned lower priority while the migration proceeds.
The runtime evidence of dependency usage allows teams to triage better by directing limited engineering capacity toward services with confirmed exposure.

\section{Threats to Validity}\label{sec:threats}
We now discuss the threats to validity of our experiments.
First, dependencies that contain exclusively annotations or interfaces are loaded by the JVM at runtime but their methods are never invoked.
Since \tool tracks method execution, these dependencies are not included in the reported set.
This may lead to an underestimation of the total number of runtime dependencies, even though, by design, \tool targets dependencies whose code is actually executed.
\revision{Second, a threat arises when the runtime classpath is expanded beyond the build-time dependency set.
In some deployment environments, such as enterprise application servers, additional JARs are contributed to the classpath at runtime.
For example, in Tomcat deployments, JDBC drivers are commonly placed in a shared library directory and shared across all deployed applications, rather than bundled within individual application artifacts.
\tool embeds metadata only for dependencies resolved at build time and bundled into the uber-JAR.
Dependencies injected at runtime in this manner are absent from the Introspector's output even if their code is actively executed.
Classport's guarantees therefore apply only to the build-time dependencies.}
\revision{Third, Classport's bytecode modification breaks bitwise identity with the original artifacts published on Maven Central, making hash-based verification against upstream checksums impossible for the embedded classes.
The same limitation applies to any uber-JAR produced by the Maven Shade Plugin, since shading itself merges and relocates class files~\cite{dietrich_security_2024,schott2025bytecodecentricdetectionknowntobevulnerabledependencies}.}
Fourth, we use the test suites of study subjects as workloads for RQ2.
Test suites may not fully represent production workloads.
They may exercise some code paths more than others and may not cover all runtime scenarios.
Thus, it is possible that the set of runtime dependencies reported by \tool is smaller than the set of dependencies actually used in production, but this is a limitation of the test suite, not of \tool.

\section{Related Work}\label{related}
In this section, we position our contribution with respect to related work on embedding dependency metadata into executables and identifying runtime dependencies in Java.

\subsection{Embedding Information into Executables and Binaries}
Boucher et al.~\cite{boucher2023automaticmaterials} propose a novel approach to identifying software dependencies, introducing ABOM, Automatic Bill Of Material. 
The main idea of ABOM is to query an executable for its dependencies to facilitate vulnerability detection.
ABOM automatically embeds hashes of dependencies into compiled binaries during the build process. These hashes are stored efficiently in a probabilistic data structure called Compressed Bloom Filters.
It allows for rapid querying to detect the presence of a specific dependency and hence to identify the vulnerabilities associated with it.
However, since the data structure is probabilistic in nature, it can report the presence of a dependency even if it is not actually present.
Our approach cannot generate false positives because it directly relies on Maven and a Maven build success implies that all dependencies for the project are resolved.

Seshadi et al.~\cite{seshadri2024omniborautomaticverifiableartifact} present OmniBOR, which embeds a content-based identifier called gitoid into executables, including Java class files, to link them to an external Artifact Dependency Graph built during compilation. This approach requires build-time instrumentation and external storage. In contrast, our method embeds Maven GAV metadata directly into class files, allowing immediate runtime access via a Java agent without relying on external resolution. Our technique is dedicated to Java and uniquely supports lightweight, runtime introspection of runtime dependencies.

Quatch et al.~\cite{quach2018debloating} propose a solution that combines static and dynamic analysis to identify and remove the unused code during the program loading process. Their framework generates function-level dependency graphs and embeds this information into an optional section of the ELF binary. This embedded dependency information is then used by a piece-wise loader at runtime to identify and remove unused code from the program's memory.
Our approach is more general since it brings dependency information from the build-time context to the runtime one and is not limited to debloating.

\subsection{Runtime Analysis and Dependency Identification}
This section reports work on runtime identification of dependencies.

Ponta et al.~\cite{ponta2020detection} present a novel code-centric and usage-based method for detecting, assessing, and mitigating vulnerabilities in open-source software dependencies.
This work builds upon their previous approaches~\cite{plate2015impactassessment, ponta2018beyondmetadata}, and is implemented in the open-source tool Eclipse Steady~\cite{eclipseSteady}.
They use a dynamic call graph of dependencies to check the reachability of the vulnerable code.
They instrument the target Java application and search if, during the execution, it reaches the vulnerable code, i.e., if it uses vulnerable methods.
The difference with our approach is that they focus on the vulnerable dependencies and not on an overall identification of the actually executed dependencies.

Soto-Valero et al.~\cite{soto2023coveragebaseddebloating} propose a tool to detect bloated code in Java and to remove it by transforming the bytecode of the compiled project.
They use four code coverage tools along with probes to identify methods that are not executed during runtime.
If no methods in any class are executed within a dependency, the dependency is considered bloated. 
They only work at testing time.
However, we embed dependency information directly in the class file in JARs, making this information available at production time as well.

Amusuo et al.~\cite{amusuo2024ztdjavamitigatingsoftwaresupply} implement a tool that is able to create a policy file with a mapping of required permissions for every dependency.
The policies are then enforced at runtime.
To create a policy file, they infer dependency namespaces by assuming that all classes within a JAR share a common root directory path, which is then mapped to the Maven Group and Artifact ID.
In contrast, we directly retrieve dependency information at runtime from metadata embedded in the class files, avoiding fragile assumptions based on package name heuristics.

Ortin et al.~\cite{ORTIN2025102250} present a tool for visualizing and comparing runtime object structures on the Java platform using reflection.
While the concept if \emph{introspection} appears in both works, the scope differs fundamentally: their work operates at the object level, inspecting fields and data structures of individual instances, whereas our approach operates at the dependency level, identifying and exposing which declared library are actually loaded and executed at runtime.

\section{Future Work}\label{sec:future}
\revision{We now discuss four directions for future work.}

\revision{First, the annotation-injection step has not yet been validated across a broad range of build configurations.
A Maven build plugin executing after the \embedder may modify or strip class files, silently removing the embedded metadata.
Batik, one candidate project, was excluded from the study because the embedder failed to inject annotations across all packages due to its reactor build configuration.
Whether such build-configuration conflicts arise more broadly, and how to choose the injection point to avoid them, warrants further investigation.}

\revision{Second, we plan to harden \tool against tampering by malicious Java agents.
Since dependency metadata is embedded as annotations, it can be modified at runtime by agents that intercept class loading.
To address this, we propose signing the class files, including dependency annotations, at build time and verifying the signature at load time.
This would ensure the integrity of dependency metadata at runtime.}

\revision{Third, while our evaluation focuses on Maven, extending \tool to other build systems such as Gradle is an engineering effort.
The \embedder and the \retriever operate on standard Java class files and are build-system-agnostic, so support would require only an equivalent Gradle plugin.}

\revision{Fourth, human aspect research can study whether application developers are willing to integrate \tool into their build and release pipelines.}

\section{Conclusion}\label{sec:conclusion}
In this paper, we proposed \tool, a novel approach to introspect runtime dependencies in Java.
\tool embeds dependency information into Java binary artifacts and exposes this information at runtime, during execution.
We demonstrated its effectiveness by augmenting \numberofprojects real-world, open-source Maven Java projects with runtime dependency introspection.
For all of them, it is possible to fully embed dependency information and to correctly identify the executed dependencies at runtime, with negligible or low overhead.

\section*{Acknowledgements}
This work was partially supported by the WASP program funded by Knut and Alice Wallenberg Foundation, and by the Swedish Foundation for Strategic Research (SSF). 
This work was partially supported by the SERICS project (PE00000014) under the MUR National Recovery and Resilience Plan, funded by the European Union - NextGenerationEU.
The authors would like to thank Deepika Tiwari for her contributions to the project.

\bibliographystyle{elsarticle-num}
\bibliography{bib.bib}

\end{document}